\documentclass[aps,pre,reprint,superscriptaddress]{revtex4-1}

\usepackage{amssymb}
\usepackage{bm}
\usepackage{graphicx}
\usepackage{amsmath}
\usepackage{amssymb}
\usepackage{gensymb}

\begin{document}


\title{Unsteady flows and inhomogeneous packing in damp granular heap flows}

\author{Hongyi Xiao}
\affiliation{Department of Mechanical Engineering, Northwestern University, Evanston, Illinois 60208, USA}

\author{John Hruska}
\affiliation{Department of Mechanical Engineering, Northwestern University, Evanston, Illinois 60208, USA}

\author{Julio M. Ottino}
\affiliation{Department of Mechanical Engineering, Northwestern University, Evanston, Illinois 60208, USA}
\affiliation{Department of Chemical and Biological Engineering, Northwestern University, Evanston, Illinois 60208, USA}
\affiliation{The Northwestern Institute on Complex Systems (NICO), Northwestern University, Evanston, Illinois 60208, USA}

\author{Richard M. Lueptow}
\affiliation{Department of Mechanical Engineering, Northwestern University, Evanston, Illinois 60208, USA}
\affiliation{The Northwestern Institute on Complex Systems (NICO), Northwestern University, Evanston, Illinois 60208, USA}

\author{Paul B. Umbanhowar}
\email{umbanhowar@northwestern.edu }
\affiliation{Department of Mechanical Engineering, Northwestern University, Evanston, Illinois 60208, USA}


\date{\today}

\begin{abstract}
We experimentally study the transition from steady flow to unsteady flow in a quasi-2D granular heap when small amounts of water are added to monodisperse glass spheres. Particles flow uniformly down both sides of the heap for low water content, but unsteady flow occurs as the water content increases. The unsteady flow mode consists of a non-depositing downslope avalanche and an upslope propagating granular jump. The transition from steady to unsteady flow occurs when the slope exceeds a critical angle as a result of water-induced cohesion. Under unsteady flow conditions, the deposited heap consists of loosely packed and densely packed layers, the formation of which is closely related to the unsteady flow dynamics.  

\end{abstract}


\maketitle

\section{Introduction}
\par Flows of wet granular materials are common in nature and industry. By increasing the liquid content, the behavior of granular materials can change dramatically from cohesionless dry particles all the way to slurries and suspensions~\cite{strauch2012wet,kudrolli2008granular,mitarai2006wet,herminghaus2005dynamics}. In this study, we focus on the first step of this transition where the particles are damp (slightly wet) and gravity-driven flow occurs. Many industrial and geophysical scenarios fall into this regime, for example, when granular materials exposed to environmental humidity become damp. Previous studies have shown that high relative humidity can significantly influence the flowability and slope stability of the materials~\cite{bocquet1998moisture,bertho2004influence,gomez2015characterization,fraysse1999humidity}. Also, a small amount of liquid can be mixed with particles intentionally to reduce segregation~\cite{samadani2000segregation,li2003controlling,liu2013effect,liao2010segregation} or suppress airborne dust~\cite{faschingleitner2011evaluation}. However, the flow behaviors of damp granular materials differ from that of dry materials. Thus, it is important to understand the influence of small quantities of added liquid on the flow. 

\par Gravity-driven free surface flows of wet granular materials exhibit unique behaviors that are relatively unexplored compared to those that occur in dry and cohesionless materials~\cite{mitarai2006wet}. The most significant difference between wet and dry granular materials is the angle of repose. An inclined bed remains static below a critical angle, $\theta_s$, defined as the maximum angle of repose~\cite{grasselli1997angles}. Previous studies have shown that for wet granular materials, $\theta_s$ depends on the liquid content, $W_c$, defined here as the volume fraction of the added liquid in the total packed volume of the particles~\cite{hornbaker1997keeps,tegzes2002avalanche,albert1997maximum,nowak2005maximum,samadani2000segregation,samadani2001angle}. At extremely low $W_c$, liquid is trapped in valleys between the asperities of particle surfaces~\cite{kohonen2004capillary,fournier2005mechanical,mason1999critical,halsey1998sandcastles}, and tiny liquid bridges may form between the asperities of two contacting particles, which introduces weak cohesion~\cite{halsey1998sandcastles,mason1999critical}. Under damp (or slightly wet) conditions, where $W_c$ is typically on the order of $10^{-4}$ to $10^{-3}$, liquid bridges can form between contacting particles due to capillary forces, which introduces cohesion~\cite{willett2000capillary,scheel2008morphological}.
The number of liquid bridges per particle, $N$, increases with increasing $W_c$ until it reaches a maximum value of approximately six bridges per particle for mono-sized spheres~\cite{scheel2008morphological}. As a result of increasing $N$, stronger cohesion between particles causes $\theta_s$ to rise significantly~\cite{hornbaker1997keeps,tegzes2002avalanche,albert1997maximum,nowak2005maximum,samadani2000segregation,samadani2001angle}. Mechanical properties of the damp material, such as yield stress and tensile strength, also increase as $N$ increases~\cite{scheel2008morphological,fournier2005mechanical}. In addition, increasing $W_c$ can cause the free surface of the flowing material to become rough, and flow instabilities may occur~\cite{samadani2000segregation,tegzes2002avalanche,tegzes2003development}, as discussed later. Further increasing $W_c$ beyond where $N=6$ leads to another regime where liquid bridges merge and form more complex structures~\cite{scheel2008morphological}, so that the angle of repose and other mechanical properties become insensitive to $W_c$~\cite{fournier2005mechanical}, which is not the focus of this study.

\par Here, we study slightly wet granular flows during heap formation, which occurs widely in both geophysical and industrial systems~\cite{khakhar2001surface,fan2012stratification,fan2013kinematics,xiao2016density,xiao2017controlling,xiao2017transient,samadani2000segregation,samadani2001angle}. For quasi-two-dimensional (quasi-2D) bounded heap flow of dry spheres~\cite{khakhar2001surface,fan2012stratification,fan2013kinematics,xiao2016density, grasselli1999shapes}, particles are fed onto the heap and flow downslope in a relatively thin flowing layer along the surface of the previously deposited particles, which form a static bed. The free surface of the flowing layer is inclined at an angle, $\theta$, which is usually slightly larger than $\theta_s$ for dry flows. For a sufficiently large and steady feed rate, the free surface rises uniformly, and particles are uniformly deposited from the flowing layer to the static bed~\cite{fan2013kinematics,fan2012stratification,khakhar2001surface}. When the particles are size bidisperse, with the smaller species being smooth and spherical and the larger species being rough and non-spherical, a second flow mode occurs and stratification of the two species can be observed~\cite{makse1997spontaneous,makse1998dynamics,cizeau1999mechanisms,grasselli1998experimental,gray1997pattern,gray2009segregation}. This flow mode is periodic with each period containing a downslope avalanche followed by a granular traveling jump that propagates upslope~\cite{makse1998dynamics}. (The jump is also referred to as a kink, a granular bore, or a shock~\cite{edwards2016size,faug2015standing,mejean2017general,gray1998particle,savage1979gravity}.) This flow mode is triggered by the interplay of particle size segregation~\cite{ottino2000mixing} and the difference in the angle of repose between the two species~\cite{cizeau1999mechanisms,grasselli1998experimental}, which is also referred as the ``segregation mobility feedback''~\cite{gray2009segregation}. Note that stratification due to size segregation can also occur for smooth spherical particles, but the underlying layer formation mechanism is different~\cite{fan2012stratification,xiao2017controlling}.

\par In three-dimensional (3D) heap formation, the flow is often not axisymmetric and unsteady flow can occur. For example, pouring certain types of sand onto a 3D heap can result in unsteady flow that revolves around the feed zone~\cite{altshuler2003sandpile,altshuler2008revolving}. The unsteady flow may be related to the segregation mobility feedback~\cite{altshuler2008revolving}, even after the particles were sifted to a relatively narrow size distribution~\cite{altshuler2003sandpile}. However, since some of these experiments were conducted at relative humidities between 60\% and 90\%~\cite{altshuler2003sandpile}, there is a possibility that water condensed on the small diameter (~0.1 mm) particles used in the study contributed to the unsteady flow~\cite{fraysse1999humidity}. For example, in experiments using damp mono-sized spheres with $W_c=4\times10^{-4}$ in rotating circular tumblers, unsteady flow occurs which consists of a downslope front followed by a second front traveling upslope~\cite{tegzes2002avalanche,tegzes2003development}, similar to that due to particle size and shape differences. This suggests that using particles differing in size and shape may not be the only way to trigger unsteady flow, and it is possible that unsteady flows can also occur for slightly wet particles in heap flows.

  \begin{figure}[t]
	
	\centerline{\includegraphics[width=3.0 in]{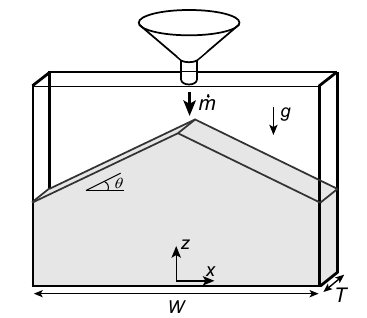}}
	
	\caption{\label{Fig1} Schematic of the experimental quasi-2D bounded heap (not to scale).}
	
\end{figure}

\par In this study, we experimentally investigate the transition from steady to unsteady heap flow of damp (slightly wetted) mono-sized spheres in a quasi-2D center-filled bounded heap geometry. Section~\ref{sec2} describes the experimental methods. Section~\ref{sec3} presents the results showing the transition from steady flow with uniform deposition to unsteady and asymmetric flow. Section~\ref{sec4} discusses the formation of the inhomogeneous packing for particles deposited on the heap. Section~\ref{sec5} presents the conclusions.

\section{Experimental methods}
\label{sec2}

\par The apparatus (Fig.~\ref{Fig1}) for the quasi-2D bounded heap experiments consist of a glass front plate and an aluminum back plate, separated by acrylic bars which form the bounding walls and the bottom. The width of the heap, $W$, is 38\,cm and the gap thickness, $T$, is 1.27\,cm. A metal funnel placed above the center of the silo feeds the particles. The outlet cross section of the funnel is shaped into a 1.1\,$\times$ 1.02\,cm$^2$ rectangle to fit the quasi-2D silo. Four sizes of glass spheres with density $\rho=2.62$\,g/cm$^3$ (Ceroglass Technologies Inc., TN, USA) are used with diameters $d$ of 0.63$\pm$0.07\,mm, 0.53$\pm$0.04\,mm, 0.35$\pm$0.05\,mm, and 0.20$\pm$0.03\,mm. The interstitial liquid is distilled water, with nominal density $\rho_w=$1\,g/cm$^3$ and surface tension $\gamma=74$\,dynes/cm.


\par For each experiment, particles are first dried in an oven at 90\,$^\circ$C for at least half an hour and then cooled in air to room temperature (21\,$^\circ$C). A volume of water, $V_w$, is mixed into an as-poured bulk volume of particles, $V_p$, ($V_p=1$\,L) in a clean glass beaker to obtain wet granular material with water content $W_c=V_w/V_p$. Experiments begin by pouring the mixture into the funnel within one minute after the mixtures are prepared to minimize evaporation. Water lost to the beaker and funnel surfaces can be neglected, as their surface areas are over 100 times smaller than the total surface area of all the particles for the largest diameter particles used here. Furthermore, water does not drain from the particles due to gravity, as the particle diameters are much smaller than the capillary length of water, $l_c=\sqrt{\gamma/\rho_w g}\approx$ 2.7\,mm, where $g$ is the acceleration due to gravity.

  \begin{figure}[b]
	\centerline{\includegraphics[width=3.7 in]{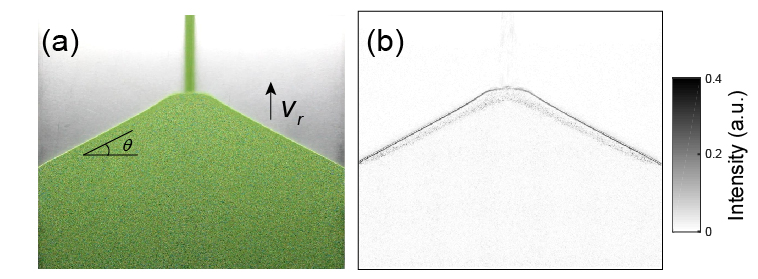}}
	
	\caption{\label{Fig2} (a) Image of rising heap with uniform deposition, and (b) image differencing result with 0.2\,s delay of steady heap flow. $d=0.63$\,mm, $W_c=0$, and $\dot{m}=64$\,g/s.}
  \end{figure}

\begin{figure*}[t]
	\centerline{\includegraphics[width=6.5 in]{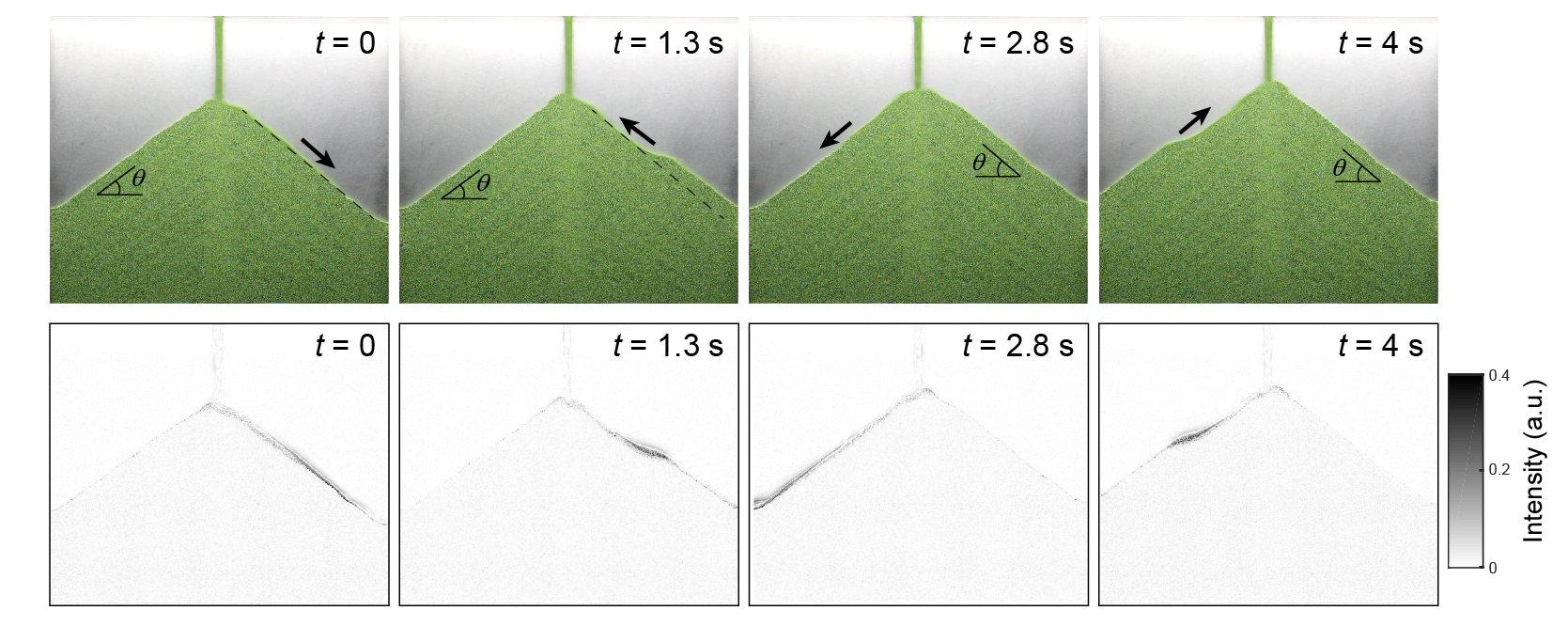}}
	
	\caption{\label{Fig3} Images (top row) and image differencing results with 0.2\,s delay (bottom row) of unsteady heap flow at different times with $d=0.63$\,mm, $W_c=0.8\times 10^{-3}$, and $\dot{m}=35$\,g/s showing an oscillatory heap instability (alternating flow). The dashed reference lines are at the same location in the $t=0$ and $t=1.3$\,s images.}
	
\end{figure*}

\par We perform experiments varying $W_c$ systematically from 0 to 1$\times$10$^{-3}$ for all particle sizes. The mass feed rate from the funnel $\dot{m}$ remains constant as the funnel empties, but decreases with increasing $W_c$, from $\dot{m}=64$\,g/s at $W_c=0$ to approximately $\dot{m}=30$\,g/s at $W_c=1\times 10^{-3}$ for the particle sizes examined. Results from additional experiments studying the influence of the feed rate are reported later in the paper.

\par A digital camera (EOS Rebel T6, Canon Inc., Japan) records videos of the experiments at 30\,frames/s with a spatial resolution of about 0.4\,mm, which is comparable to one particle diameter. The videos allow us to measure the instantaneous surface height profile $h(x,t)$ by examining the change of image intensity in each column of the image~\cite{xiao2017transient}, where $x$ is the coordinate in the horizontal direction, and $z$ is the vertical direction with the origin at the center of the heap base, see Fig.~\ref{Fig1}. In addition, a high speed camera (Flea3, Point Grey Research Inc., Canada) records videos for specific smaller regions of the flow at 200 frames/s with a spatial resolution of 0.1\,mm.

  \begin{figure}[t!]
	
	\centerline{\includegraphics[width=3.8 in]{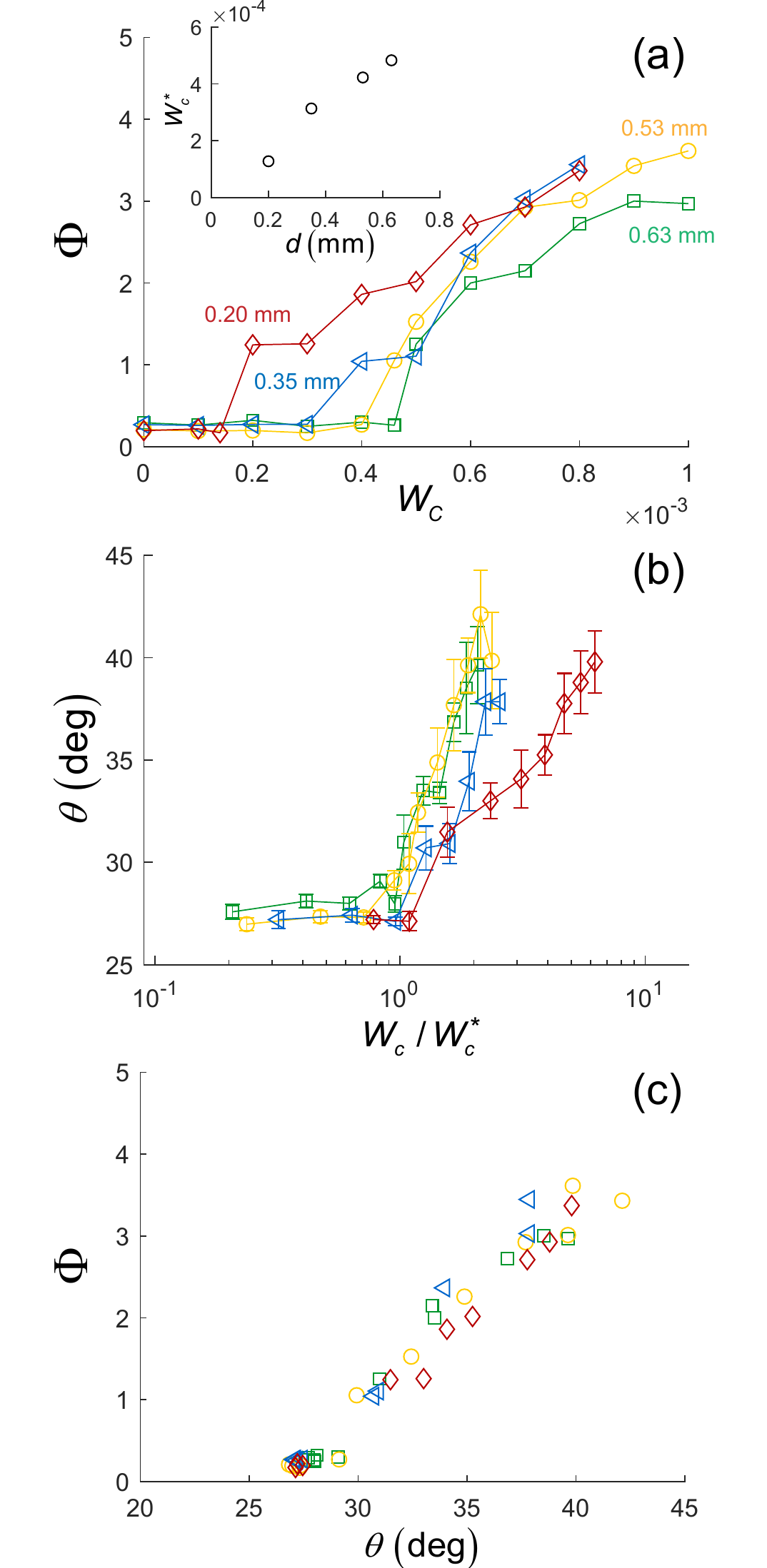}}
	
	\caption{\label{Fig4} Results of the parametric study varying $W_c$ and $d$. (a) Change of the flow unsteadiness index $\Phi$ with $W_c$. Inset: $W_c^*$ vs. $d$. (b) Surface angle $\theta$ vs. $W_c/W_c^*$. (c) Change of $\Phi$ with $\theta$. Particle sizes: 0.20\,mm (red diamonds), 0.35\,mm (blue triangles), 0.53\,mm (yellow circles), and 0.63\,mm (green squares).}
	
\end{figure}

\section{Transition to unsteady flows}
\label{sec3}

  \begin{figure*}[t]
	
	\centerline{\includegraphics[width=5.8 in]{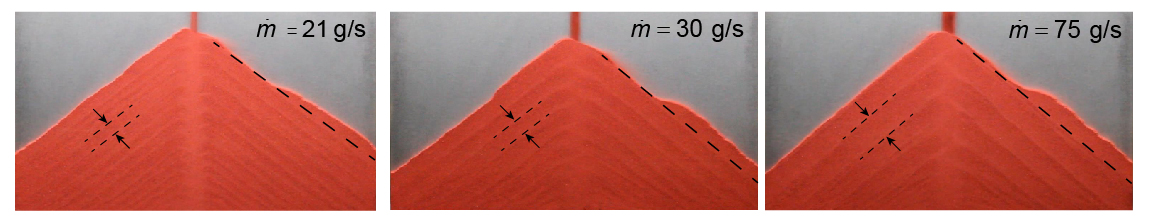}}
	\caption{\label{Fig5} Images of heap flow with $d=0.20$\,mm and $W_c=0.7\times10^{-3}$ at different feed rates showing strong dependence of the jump height slope on the feed rate but weak dependence of the heap surface angle on the feed rate. Surface angle $\theta=37.5^\circ$ for $\dot{m}=21$\,g/s, $\theta=38.2^\circ$ for $\dot{m}=30$\,g/s, and $\theta=38.8^\circ$ for $\dot{m}=75$\,g/s. Deposited layer thickness is indicated by the arrows.}
  \end{figure*}

\par We first characterize the two flow modes observed at different $W_c$. With zero or small $W_c$, steady heap flow occurs with uniform deposition on both sides of the heap, as described in previous studies of dry bounded granular heap flows~\cite{khakhar2001surface,fan2012stratification,fan2013kinematics}. An example with $d=0.63$\,mm and $W_c=0$ is shown in Fig.~\ref{Fig2}(a), where the free surface is symmetric about the center and inclined at $\theta=27.7^\circ$ during the flow. As more material is fed into the system, the free surface rises uniformly at a constant rise velocity, $v_r=\dot{m}/(\phi \rho WT)$, where $\phi=0.60$ is the packing density of the heap for the dry particles used here. To visualize the flowing layer, we measure the image intensity difference between two frames with a time increment of 0.2\,s. In this way, flowing regions have a large image difference (dark) while non-flowing regions have a negligible image difference (light). The result [e.g.,~Fig.~\ref{Fig2}(b)] shows that flow occurs in a thin layer corresponding to the gray region near the free surface. The thin dark layer on the free surface results from the surface rising uniformly.

\par At larger $W_c$, the flow is unsteady and asymmetric about the center of the heap. A time series of images for the flow with $d=0.63$\,mm and $W_c=0.8\times 10^{-3}$ is shown in Fig.~\ref{Fig3}. At $t=0$, flow occurs only on the right side of the heap in the form of an avalanche propagating downstream. After the avalanche front reaches the bounding wall, an upstream traveling jump~\cite{savage1979gravity,faug2015standing,mejean2017general} forms ($t=1.3$\,s). After the jump reaches the feed zone near the center, it directs the feed stream toward the left side of the heap, which triggers a downslope avalanche on the left side ($t=2.8$\,s). In the meantime, the right side becomes static, as the image difference shows. When the front of the downslope avalanche reaches the left bounding wall, an upstream traveling jump forms at that bounding wall ($t=4$\,s). In this way, the flow is periodic and continues to alternate between the two sides of the heap. 

\par The downslope avalanche and upslope traveling jump are similar to the unsteady and periodic flows observed in studies of spontaneous stratification due to ``segregation mobility feedback''~\cite{makse1997spontaneous,makse1998dynamics,cizeau1999mechanisms,grasselli1998experimental,gray1997pattern,gray2009segregation}. However, since monodisperse spheres do not segregate, a different mechanism must govern the transition. In addition to the unsteady flow mode for $W_c=0.8\times 10^{-3}$, the surface incline of $\theta=33.4^\circ$ is steeper than that with no water ($\theta=27.7^\circ$). For unsteady alternating flow, the angle $\theta$ is measured as the angle of the free surface on the static side. This is the angle formed by the upslope traveling jump in the previous period, and is also the angle of the slope on which the next downslope avalanche will propagate. Another significant difference with the dry flow case is that slightly darker and lighter layers are evident in the deposited heap (observable in Fig.~\ref{Fig3}). The lighter layers are densely packed, and the darker layers are loosely packed. The loosely packed layers have more voids which make them look darker when the apparatus is lit from above. The formation of these layers is discussed in detail in Section~\ref{sec4}.

\par To quantify the transition from steady to unsteady flow, we first define an unsteadiness index, $\Phi$:

\begin{equation}
\Phi=\left<\left[ \frac{1}{W}\int_{-W/2}^{W/2}\frac{\left(v_r(x,t)-v_{r0}\right)^2}{v_{r0}^2}dx\right]^{1/2} \right>.
\label{Eq1}
\end{equation}

\noindent In this relation, $v_r(x,t)=dh(x,t)/dt$ is the instantaneous local surface rise velocity, $v_{r0}=\dot{m}/\phi \rho WT$ is the average rise velocity, and $\left<~\right>$ denotes a temporal average measured from when the heap base first spans the entire width of the apparatus to when the feed is stopped. Thus, $\Phi$ is essentially a spatial and temporal average of the local deviation from steady flow. We exclude from the average the 5\,cm wide feed zone in the center and the 2.5\,cm wide regions adjacent to each downslope bounding wall to reduce the influence of bouncing particles on $\Phi$. For steady flows (Fig.~\ref{Fig2}), $v_r(x,t)\approx v_{r0}$ (with small fluctuations), so that $\Phi$ is close to zero. For unsteady flows, static regions with $v_r(x,t)=0$ yield a local deviation of 1, while $v_r(x,t)$ for the traveling fronts of the downslope avalanche and the upslope traveling jump are much greater than $v_{r0}$, and result in local fluctuations greater than 1. Thus, on average, the unsteadiness index $\Phi$ is typically greater than 0.5 for unsteady flows, and we define a transition water content, $W_c^*$, which corresponds to the water content when $\Phi=0.5$ for a particular particle size.

\par Figure~\ref{Fig4}(a) shows how $\Phi$ varies with liquid content for the four particle sizes considered. For each particle size, $\Phi$ is close to zero at $W_c=0$ and remains near zero at small $W_c$, indicating that the flow is steady. As $W_c$ is further increased, $\Phi$ starts to increase, indicating unsteady flow. When $W_c$ is only slightly larger than the transition water content, $W_{c}^*$, the propagating fronts are less distinct than those shown in Fig.~\ref{Fig3} and sometimes do not propagate the entire length of the slope. Thus, the deviation of $v_r(x,t)$ from $v_{r0}$ is relatively small resulting in $\Phi$ slightly above 1. When $W_c$ is further increased, the propagating fronts are sharper which results in larger $\Phi$. Near $W_c=1\times$10$^{-3}$, the free surface becomes rough such that more localized flows (i.e., not spanning the entire slope) and avalanches occur~\cite{tegzes2002avalanche,tegzes2003development,samadani2000segregation}, which results in a plateau in $\Phi$ for the larger particles ($d=0.53$\,mm and $d=0.63$\,mm). Note that data for $W_c>0.8\times10^{-3}$ for the smaller particles are not available because the feed funnel jams at and above this water content. Figure~\ref{Fig4}(a) also indicates that the transition to unsteady flow occurs at smaller water contents for smaller particles. To further demonstrate this trend, the transition water content $W_c^*$ is plotted versus particle diameter $d$ in the inset of Fig.~\ref{Fig4}(a), which shows that $W_c^*$ increases with $d$. Note that a linear interpolation is applied to calculate $W_c^*$ at $\Phi=0.5$ using two neighboring data points.

\par The transition from steady to unsteady flow is likely related to the ratio between water-induced cohesion and particle weight. Steady flow occurs at zero or small cohesion, while unsteady flow occurs when cohesion is significant compared to the particle weight. The liquid bridge force between two particles can often be approximated as $\pi\gamma d$~\cite{kudrolli2008granular,willett2000capillary,herminghaus2005dynamics}, while the particle weight is $\frac{\pi}{6}\rho g d^3$. Thus, a Bond number that characterizes the ratio between liquid bridge force and particle weight can be defined as, $Bo=6\gamma/\rho gd^2$~\cite{willett2000capillary,zhu2013effects,liao2016effect}. The Bond number scales with $d^{-2}$, indicating that liquid bridge force can dominate for smaller particles. However, the Bond number alone is not adequate to explain the increase of $\phi$ with $W_c$ in Fig.~\ref{Fig4}(a) since $Bo$ is independent of $W_c$. This is because that the liquid bridge force is relatively insensitive to the bridge volume~\cite{willett2000capillary,kudrolli2008granular,herminghaus2005dynamics}. 

\par To explain the dependence of $\Phi$ on $W_c$, two possible scenarios can be considered. The first one is related to particle roughness and predicts that the transition from steady to unsteady flow is determined by the initial formation of liquid bridges, which occurs when the valleys between surface asperities are filled with liquid~\cite{kohonen2004capillary,halsey1998sandcastles,mason1999critical}. This occurs at $W_{c,as}=6\alpha\phi\delta/d$~\cite{kohonen2004capillary}, where $\alpha$ is the ratio of the area of valleys between asperities to the total surface area of a particle, and $\delta$ is the characteristic height of the asperities. Assuming that surface roughness does not vary with particle size, the asperity filling mechanism predicts that $W_c^*$ should decrease with increasing $d$, which is opposite to the observations shown in Fig.~\ref{Fig4}(a). In the second scenario, the transition occurs at $W_c^*>W_{c,as}$, and the controlling parameter becomes the number of liquid bridges per particle, $N$, which increases with $W_c$ from 1 to a saturating value of 6~\cite{kohonen2004capillary,scheel2008morphological,fournier2005mechanical}. In this regime, the overall force ratio for a particle is $NBo$, which increases with $W_c$, and this causes the material properties such as tensile strength and yield stress to increase with $W_c$~\cite{fournier2005mechanical}. Similarly, the increase of $NBo$ can also drive the increase of $\Phi$ with $W_c$ in Fig.~\ref{Fig4}(a). In addition, the fact that lower water content is required to trigger the flow transition for smaller particles [Fig.~\ref{Fig4}(a) inset] can be explained: to reach the same overall force ratio required for the flow transition, fewer liquid bridges ($N$) are needed for smaller particles, which in turn lowers $W_c^*$. Note that the exact functional form for the relation between $W_c^*$ and $d$ is difficult to specify, as many complications likely come into play, such as the possible dependence of $N$ on $d$~\cite{kohonen2004capillary}.

\par Having identified the water content as a critical factor for the flow transition, we further discuss the mechanism for the transition and focus on another important property that is significantly influenced by $W_c$: the surface angle $\theta$. Figure~\ref{Fig4}(b) shows that for the particle sizes examined, $\theta$ is approximately 27$^\circ$ for $W_c$ approaching zero. It increases abruptly for $W_c/W_c^*>1$ to approximately 40$^\circ$ for the maximum water content $W_c\approx 1\times$10$^{-3}$, which agrees qualitatively with previous studies~\cite{hornbaker1997keeps,tegzes2002avalanche,albert1997maximum,nowak2005maximum,samadani2000segregation,samadani2001angle}. Again, this increase is most likely related to the increase of $NBo$: stronger cohesion requires steeper angle for gravity-driven flow to occur. Note that for $W_c/W_c^*<1$, the increase in $\theta$ is small compared to the increase for $W_c/W_c*>1$, as shown in Fig.~\ref{Fig4}(b). This is not unexpected because the surface slope in steady and unsteady flows develops under different flow dynamics and cohesion can play different roles in the formation of the slope in the two flow modes.

\par Figure~\ref{Fig4}(c) shows the relation between $\Phi$ and $\theta$, which indicates a flow transition at $\theta\approx30^\circ$ for all particles sizes. For $\theta<30^\circ$, the flow is steady ($\Phi$ close to zero) and the surface rises uniformly (Fig.~\ref{Fig2}), indicating uniform particle deposition on the entire heap. For $\theta>30^\circ$, the flow becomes unsteady ($\Phi>1$), as shown in Fig.~\ref{Fig3}, and no deposition occurs during the downslope avalanche ($t=0$ and $t=2.8$\,s in Fig.~\ref{Fig3}). This is evident by comparing the material above the dashed reference lines in Fig.~\ref{Fig3} at $t=0$ and $t=1.3$\,s, where the surface upslope from the traveling jump does not rise (i.e., there is no deposition in this region). Thus, the downstream flow experiences a transition from steady deposition to unsteady deposition (only during the upslope traveling jump on one side of the heap) as $\theta$ increases beyond $30^\circ$ due to the water-induced cohesion. 

\par This transition is reminiscent of the concept of a ``neutral angle,'' $\theta_n$, that has been discussed in a number of studies focusing on material exchange between the flowing layer and the underlying static bed~\cite{bouchaud1994model,boutreux1998surface,trinh2017erosion}. These studies propose that deposition of particles on the static bed during the downstream flow is only possible when $\theta<\theta_n$, where $\theta_n=30^\circ$ in this study. For the unsteady flows with $\theta>30^\circ$, deposition is only possible via the upslope traveling jump, where particles flow along the slope to the traveling jump and are deposited on the face of the jump. Previous studies of dry granular flows in a chute indicate that a minimum base incline angle and a downstream obstacle are required for the formation of the jump~\cite{savage1979gravity,faug2015standing,mejean2017general}, as mentioned previously. Here, increased $\theta$ due to cohesion and the presence of the downstream bounding walls satisfy these two conditions, respectively. Note that no upslope propagating jumps were found in experiments where the bounding walls (the obstacles) are removed. In addition, the unsteady flow mode observed here resembles the unsteady flow mode formed due to the segregation of particles with size and shape differences~\cite{makse1997spontaneous,makse1998dynamics,cizeau1999mechanisms,grasselli1998experimental,gray1997pattern,gray2009segregation}. In that case, large and rough particles segregate to the free surface, which sets a higher surface angle (possibly higher than $\theta_n$ for the small spherical particles that segregate to the interface between the flowing layer and the static bed) to induce the downslope avalanche and the upslope traveling jump~\cite{makse1998dynamics}. Thus, it appears that particle roughness in size-and-shape stratification flow and water-induced cohesion in slightly wet flow play similar roles in increasing $\theta$ beyond $\theta_n$, resulting in a similar traveling jump flow mode. 

\par Lastly, we consider the influence of the feed rate $\dot{m}$ on heap flow. As mentioned in Sec.~\ref{sec2}, the feed rate from the funnel decreases as $W_c$ increases. However, $\dot{m}$ plays a lesser role in the transition to unsteady flow than does $W_c$. Figure~\ref{Fig5} shows three cases with the same particle diameter and water content but with different funnel sizes that produce different feed rates: $\dot{m}=21$\,g/s (small funnel), $\dot{m}=30$\,g/s (medium funnel which corresponds to the data in Fig.~\ref{Fig4}), and $\dot{m}=75$\,g/s (large funnel), which is higher than the dry feed rate of the medium funnel (64\,g/s). At all three feed rates, unsteady flow occurs but the increase of $\theta$ due to $\dot{m}$ is only about 1$^\circ$, which is less significant than the increase of $\theta$ due to $W_c$ (more than 10$^\circ$), indicating that the flow mode is less sensitive to $\dot{m}$ than to $W_c$. The shape and the height of granular jumps in the related problem of dry granular chute flows are mainly determined by the flow rate and the incline angle of the chute~\cite{faug2015standing,mejean2017general}. Here, the height of the jump in slightly wet flows also increases with $\dot{m}$ as shown in Fig.~\ref{Fig5}. A consequence of varying $\dot{m}$ is that the thickness of the alternating layers deposited on the heap also varies, which will be discussed later. Quantifying the scaling and other details related to the influence of $\dot{m}$, $W_c$, and $d$ on the shape and height of the traveling jump are topics for future work.

  \begin{figure}[b]
	
	\centerline{\includegraphics[width=3.6 in]{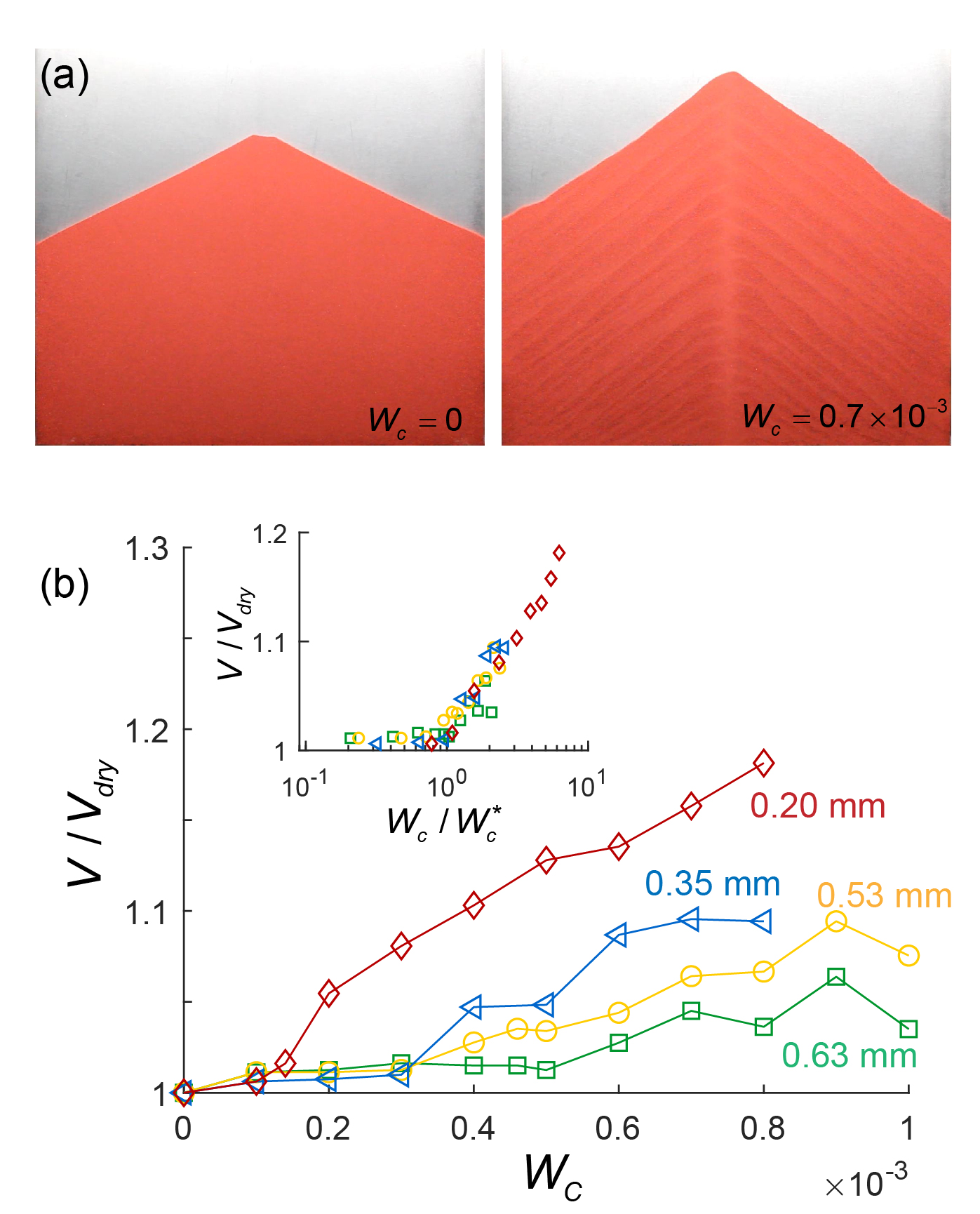}}
	
	\caption{\label{Fig6} Increase in volume of the deposited heap in slightly wet granular flows. (a) Images of the final heap for different water contents with $d=0.20$\,mm particles. (b) Volume ratio $V/V_{dry}$ vs. $W_c$ for the four particle sizes. Inset shows $V/V_{dry}$ vs. $W_c/W_c^*$.}
	
  \end{figure}

\section{Inhomogeneous packing}
\label{sec4}

\par The faint alternating lighter and darker layers evident in Figs.~\ref{Fig3}~and~\ref{Fig5} are a consequence of heterogeneous packing in the deposited heap. Previous studies have shown that the poured packing density can be less dense for granular materials with cohesive forces due to adhesion (van der Waals forces)~\cite{liu2017influence,parteli2014attractive,yang2007simulation} or liquid addition~\cite{feng1998effect,li2014similarity,yang2003numerical}. Here, we also observe that the overall packing density in the deposited heap formed by slightly wet flows is less than that for dry flows. Figure~\ref{Fig6}(a) compares heaps of $d=0.20$\,mm particles for $W_c=0$ and $W_c=0.7\times10^{-3}$. Although the dry weight of the two heaps is equal, the volume for $W_c=0.7\times10^{-3}$ is clearly larger than the volume for $W_c=0$. In addition, it is also clear that the surface incline (the dynamic angle of repose) is steeper for $W_c=0.7\times10^{-3}$ and the layering in the deposited heap only occurs with the damp unsteady flow.

\par To quantify the change in the overall packing density, we plot the volume ratio $V/V_{dry}$ versus $W_c$ in Fig.~\ref{Fig6}(b), where the total volume $V=T\int_{-W/2}^{W/2}h(x)dx$ of the deposited heap for wet flows is larger than the volume for dry flows, $V_{dry}$. For all particle sizes, $V/V_{dry}$ first increases slightly with $W_c$ indicating that the packing is marginally less dense for steady heap flow. As $W_c$ is further increased and the flow becomes unsteady, the increase in $V/V_{dry}$ with $W_c$ is much steeper. At $W_c$ near $1\times10^{-3}$, $V/V_{dry}$ plateaus except for the case of $d=0.20$\,mm particles where data for $W_c>0.8\times10^{-3}$ are not available due to funnel jamming. In addition, for the same $W_c$, the increase in the heap volume $V/V_{dry}$ is more significant for smaller particles. This is likely a result of the difference in the force ratio between water induced cohesion and particle weight for different sized particles, as discussed in Sec.~\ref{sec3}. Thus, we apply the same scaling by plotting $V/V_{dry}$ versus $W_c/W_c^*$ in the inset of Fig.~\ref{Fig6}(b), resulting in the collapse of the data except for the few cases near $W_c=1\times10^{-3}$ ($W_c/W_c^*\approx 2$ for the largest particles) where $V/V_{dry}$ plateaus. This result is similar to the results in a previous computational study where the packing density decreases when the ratio between cohesion and gravity is increased~\cite{yang2007simulation}, but that study reports no layers of differing packing densities. In addition, the inset of Fig.~\ref{Fig6}(b) also shows that the increase of $V/V_{dry}$ with $W_c/W_c^*$ transitions at $W_c/W_c^*\approx1$, which corresponds to the transition from steady to unsteady flow, indicating that the transition to unsteady flow influences the packing structure of the deposited heap.

\par As mentioned earlier, slightly lighter and darker layers occur in the deposited heap for unsteady flow, as shown in Figs.~\ref{Fig3},~\ref{Fig5},~and~\ref{Fig6}(a). Particles in the lighter layers are densely packed while particles in the darker layers are loosely packed with more voids evident, at least near the clear front wall. This inhomogeneous distribution of the packing density has not been reported in previous studies on packing of cohesive granular materials~\cite{parteli2014attractive,feng1998effect,li2014similarity,liu2017influence,yang2007simulation,yang2003numerical}. Since it is known that wet particles pack loosely and dry particles pack densely, one may wonder if the particles in the densely packed layers are dryer and {\it{vice versa}}. This can be determined using fluorescent imaging. A green fluorescent dye (Model 295-17, Cole-Parmer Instrument Co., IL, USA) is added to distilled water at a concentration of 2.3\,mg/ml, and experiments are conducted using the same protocol as with undyed water. An Ultraviolet (UV) light with wavelength 365\,nm (Model XX-15N, Spectronics Inc., NY, USA) illuminates the deposited heap, and a digital camera acquires images. A longpass filter (GG495, Thorlabs, Inc., NJ, USA) placed in front of the camera filters out visible light below a wavelength of 495\,nm, and a UV filter (Model 54-058, Edmund Scientific Inc., NJ, USA) in front of the longpass filter blocks the UV light from reaching the longpass filter and the camera. No significant influence on the flow from the fluorescent dye is evident.

  \begin{figure}[t]
	\centerline{\includegraphics[width=3.8 in]{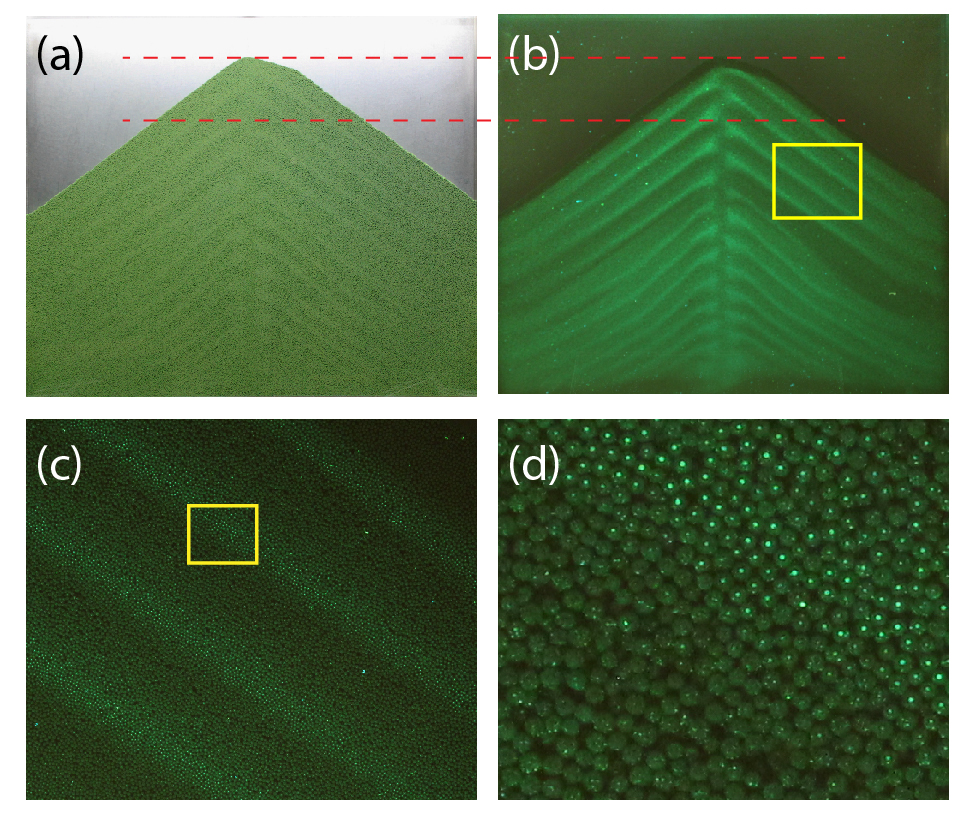}}
	
	\caption{\label{Fig7} Fluorescent light images of a heap with $W_c=0.8\times10^{-3}$, $d=0.63$\,mm, and $\dot{m}=33$\,g/s. (a) Image of the deposited heap under visible light. (b) Image of the deposited heap under UV light, and magnified images of (c) yellow box in (b), and (d) yellow box in (c). Horizontal lines in (a,b) show correspondence between features in (a) visible and (b) fluorescent images.}
	
  \end{figure}

  \begin{figure*}[t]
	\centerline{\includegraphics[width=6.9 in]{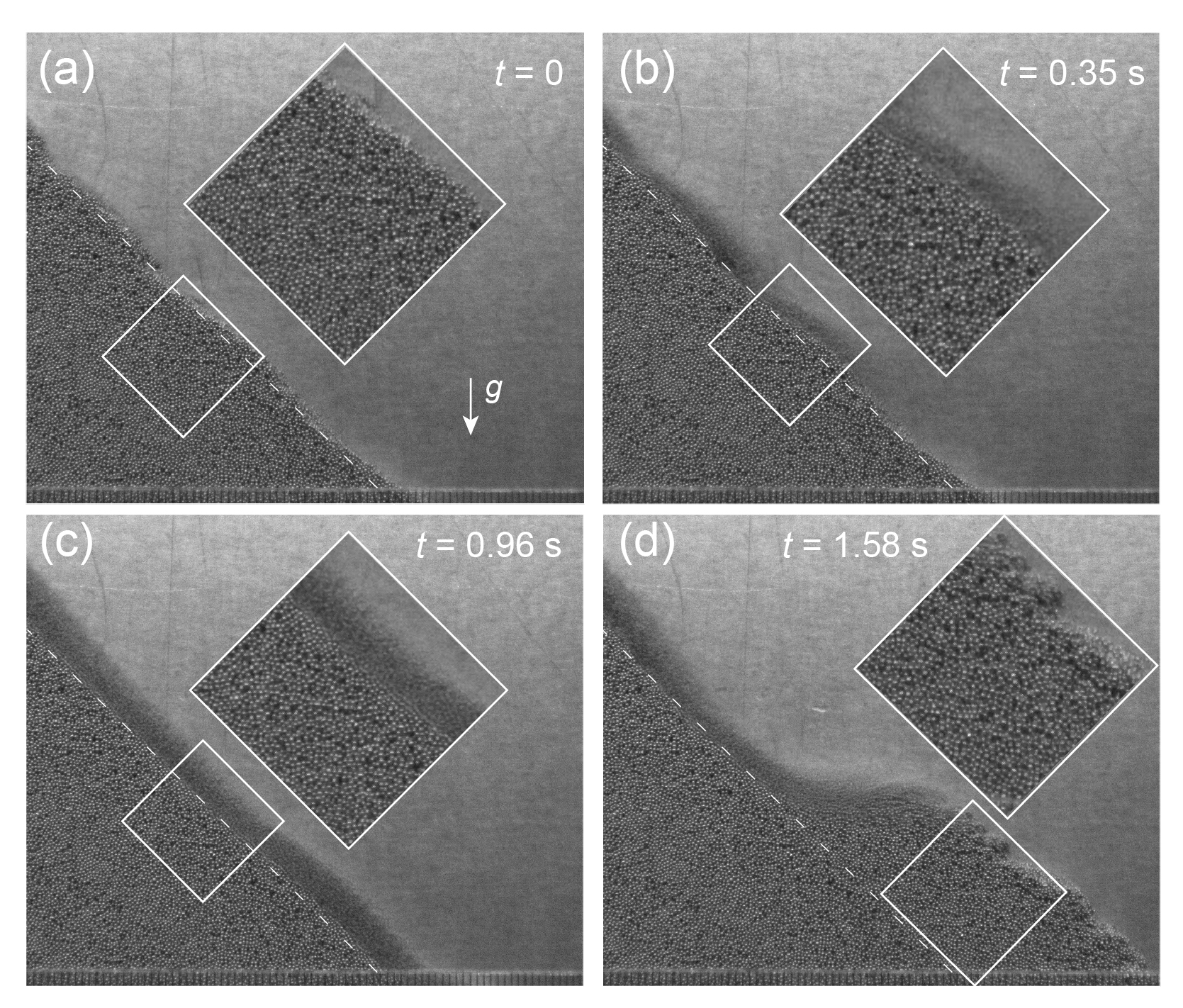}}
	\caption{\label{Fig8} Images of an 80 $\times$ 67\,mm region halfway down the right slope with $d=0.53$\,mm, $W_c=0.6\times 10^{-3}$, and $\dot{m}=37$\,g/s. Insets show magnified surface regions as indicated. (a) Static slope when flow is on the left side. (b) Downslope avalanche front enters the window. (c) Flow after the downslope avalanche front passes. (d) Flow when the upslope jump propagates into the region. Dashed white reference lines are located at the same position in all panels.}
  \end{figure*}

\par Figure~\ref{Fig7} shows an example of the fluorescent imaging results with $d=0.63$\,mm and $W_c=0.8\times10^{-3}$. Figure~\ref{Fig7}(a) shows the deposited heap under visible light, where the lighter and darker layers are visible. The corresponding image under the UV light is shown in Fig.~\ref{Fig7}(b). Note that the top layer of the free surface is dark, probably due to evaporation immediately after the heap was deposited. More importantly, there are lighter and darker layers in the fluorescent image of the heap. Examination of Figs.~\ref{Fig7}(a) and (b) indicates that the lighter and densely packed layers in Fig.~\ref{Fig7}(a) are also the lighter layers in Fig.~\ref{Fig7}(b), demonstrating that the particles within these layers are not dry. To better understand this, Fig.~\ref{Fig7}(c) shows an enlarged image of the layers under UV light, and Fig.~\ref{Fig7}(d) shows an even further magnified image of two layers. Here, in the densely packed layer [upper-right region in Fig.~\ref{Fig7}(d)], a lighter spot appears on each particle. These spots are liquid bridges formed between particles and the front glass wall. For the loosely packed layer [lower-left region in Fig.~\ref{Fig7}(d)], few liquid bridges between particles and the front glass wall are evident because the loosely packed particles make fewer contacts with the front wall. Thus, the loosely packed layers appear darker in Figs.~\ref{Fig7}(b) and (c), although particles from both the densely packed and loosely packed layers are wetted. Also, from Fig.~\ref{Fig7}(b), which shows the layering more clearly, it is evident that the layers on the left and right sides of the heap are asymmetric about the center of the heap, corresponding to the alternating unsteady flow and indicating a strong connection between the dynamics of the flow and the formation of the layers.

\par To demonstrate how the layers form during the unsteady flow, Fig.~\ref{Fig8} shows a time series of images acquired by focusing on a small region of the heap using the high speed camera. The image in each panel is an average of frames recorded over a 0.05\,s period: regions with moving particles are blurred and regions with no motion remain sharp. Figure~\ref{Fig8}(a) shows an inclined static surface halfway down the slope when the flow is on the opposite side of the heap at time $t=0$. In this image, the layer near the free surface is loosely packed, as evident by the dark voids in the close up image, and the free surface is rough, similar to the free surfaces observed in previous experiments of damp granular flows~\cite{tegzes2003development,samadani2001angle}. A dashed reference line indicating the free surface location is reproduced at the same position in the three subsequent images. 

\par Shortly after $t=0$, the flow switches to the right side, and at $t=0.35$\,s the downslope avalanche front, which is the blurred region on the upper-left portion of the surface in Fig.~\ref{Fig8}(b), enters the image. After the front passes, particles continue to flow through this image window, shown in Fig.~\ref{Fig8}(c) ($t=0.96$\,s). In this image, a thin flowing layer with a thickness of about 10$d$ is observed on the free surface, while the particles below the flowing layer remain static with a clear interface between these two regions. The interface coincides with the reference line. There is minimal deposition or erosion that occurs between the flowing layer and the static region, though the bump of particles above the reference line in the close-up image in Fig.~\ref{Fig8}(a)~and~\ref{Fig8}(b) has been eroded in Fig.~\ref{Fig8}(c). In this case, $\theta\approx 40^\circ$ because of the cohesion, so no deposition occurs during the downslope flow. Except for smoothing the surface, no erosion occurs, possibly because the cohesion also increases the yield stress in the static region~\cite{fournier2005mechanical,scheel2008morphological}. However, particles immediately below the flowing layer become densely packed evident in the close-up image in Fig.~\ref{Fig8}(c), leaving less densely packed particles below them, evident in the close up image. Apparently, collisions of particles in the flowing layer with particles in the bed smooth and compact the very top of the ``static'' region, and the interface becomes smooth. 

\par Finally, at $t=1.58$\,s, the upslope traveling jump propagates into the image window from the lower right, shown in Fig.~\ref{Fig8}(d). The jump in this case has a height of approximately 30$d$. The particles in the thin flowing layer approach the jump, climb up its face and then come to rest, thus propagating the jump upstream. Particles closer to the free surface are more loosely packed and form a rough surface, while particles deeper below the free surface are pressed together and are more densely packed. This results in a depthwise packing density gradient in the newly deposited region behind the jump. After the traveling jump passes, the slope is similar to the static slope at $t=0$ in Fig.~\ref{Fig8}(a).

\par Figure~\ref{Fig8} indicates that the dynamics of the unsteady flow at higher water content accounts for layering in the packing density. The loosely packed layers observed in Figs.~\ref{Fig3},~\ref{Fig5},~\ref{Fig7}(a),~and~\ref{Fig8}(a) are formed when the traveling jumps propagate upslope. Consequently, the thickness of these layers is set by the height of the jumps. This can be demonstrated by comparing the layering patterns in Fig.~\ref{Fig5} where the thickness of the layer in each case is indicated by the arrows. For $\dot{m}=21$\,g/s, the jump height is small and the layers are thinner than the layers for $\dot{m}=75$\,g/s, which has a higher jump. As discussed in previous studies of dry granular jumps demonstrated in chutes~\cite{mejean2017general,faug2015standing}, the shape of the jump is influenced by the base incline angle, the incoming velocity of the particles, and the thickness of the flowing layer. It is also possible that the packing density gradient in the depthwise direction varies depending on the height and the shape of the jump. However, exploring the relationship between the traveling jump and the layer packing density requires careful measurement of the flowing layer thickness, particle velocity, jump height, and the packing density distribution in the jump, which is beyond the scope of this study, but should be considered for future work.

\section{Conclusions}
\label{sec5}

\par In summary, we experimentally studied granular flows of damp sub-millimeter glass spheres in a quasi-2D bounded heap with a water content $W_c$, volume of water to volume of particles, ranging from 0 to $1\times10^{-3}$. At zero or low $W_c$, steady flow occurs simultaneously and uniformly down both sides of the heap. At higher $W_c$, the flow becomes periodic. Each period is composed of a downslope avalanche and an upslope jump on alternating sides of the heap, similar to the flow pattern in spontaneous stratification of smooth and rough particles~\cite{makse1998dynamics}. The flow is asymmetric on the two sides of the heap resembling the flow asymmetry found in 3D heap flows~\cite{altshuler2003sandpile,altshuler2008revolving}. The transition from steady to unsteady flows occurs when the surface angle is increased due to cohesion beyond a critical value which is approximately 30$^\circ$ for the case of sub-mm spherical glass particles studied here.

\par In addition to the flow mode transition, the packing density of the deposited heap for wet flow is lower than that for dry flow. It is known that cohesive particles tend to have a less dense poured packing density than non-cohesive particles~\cite{parteli2014attractive,feng1998effect,li2014similarity,liu2017influence,yang2007simulation,yang2003numerical}. However, here we show that in addition to the reduced packing density, the packing in the deposited heap of damp granular materials is inhomogeneous with densely packed and loosely packed layers occurring as a result of the unsteady flow dynamic. These layers are formed during the upslope propagation of the traveling jump. 

\par The mechanism for the wetted flow transition studied here could help in understanding unsteady flows in other flow geometries with particles slightly wetted either by adding liquid or by environmental humidity. The inhomogeneous packing could have important implications in various aspects such as slope stability, mechanical properties, thermal conductivity, and permeability of heaps in industrial and geophysical situations. Future work should focus on quantifying the formation of the traveling jump and its relation to the inhomogeneous packing distribution, as well as its implications for 3D heap formation.

\begin{acknowledgments}
This material is based upon work supported by the National Science Foundation under Grant No. CBET-1511450.
\end{acknowledgments}


%

\end{document}